# Full-Form Model-Free Adaptive Control for a Family of Multivariable System

Feilong Zhang
State Key Laboratory of Robotics, Chinese Academy of Sciences, Shenyang 110016, China

*Abstract*—This correspondence proposes a kind of model-free adaptive control (MFAC) on the basis of full-form equivalent-dynamic-linearization model (EDLM) for the multivariable nonlinear system. Compared with the current MFAC, i) this control law does not have denominator, which is stemmed from the norm of the inverse matrix and it inevitably misses the coupling relationships among the inputs and outputs (I/O) of systems. ii) the current restrictive assumption of a diagonally dominant matrix is reduced to extend its application. iii) the MFAC based on full-form EDLM is more general than the current MFAC based on partial-form and compact-form EDLM. At last, the convergence of tracking error and the BIBO stability of controlled system have been proved, which is one of the open questions in MFAC.

## I. INTRODUCTION

The developments in MFAC for single-input single-output (SISO) nonlinear systems have been witnessed recently. Hou has presented the MFAC controller on the basis of three kind of discrete-time EDLM: full-form EDLM, partial-form EDLM and compact-form EDLM [1]-[8]. The merits of MFAC are that the controller design only relies on the I/O data of systems, the model reduction, modeling error and unmolded dynamics are avoided [9]. Since pseudo-gradient (PG) vector is the only requirement for the controller design and its components are the coefficients of the EDLM. The time-varying PG is only determined by the input and output data of the system [1]-[10]. In real applications, MFAC has performed in several fields, such as: pneumatic artificial muscle [11], autonomous parking systems [12], power system or microgrids system [13]-[15] and motor control system [16]. Since the discrete simple structure of the MFAC makes it easier be applied by computers.

For multi-input multi-output (MIMO) systems, owing to the couplings among the input and output, most of conclusions of MFAC in SISO systems can't be extended to more complicated MIMO systems directly. Therefore, fewer results about MFAC for multivariable systems are available in the literature [17]. Furthermore, the stability analysis and the proof of convergence of the tracking error are still the open question for MFAC based on full-form EDLM in the MIMO system [1][9], which is relatively difficult to analyze the stability than the current MFAC based on partial-form EDLM and compact-form EDLM in MIMO system. In addition, the current MFAC methods in MIMO system change the inverse matrix into the reciprocal of matrix norm to help the stability analysis, however, this will lead to the loss of inputs and outputs coupling information. Meanwhile, current MFAC are based on a special kind of diagonally dominant matrix assumption for the pseudo Jacobian matrix (PJM), this note shows that the previous assumption turns out to be slightly more restrictive and can be extended wider.

Some major contributions of this correspondence lie as follows.
1) The full-form equivalent dynamical linearization model in MIMO systems is firstly proved, to the author's best knowledge.
2) We reduce the assumption about the PJM, and the stability of the FFDL-MFAC in MIMO systems is still guaranteed.
3) Taking the norm of the inverse matrix part of the current MFAC may lose the coupling relationships of inputs and outputs, so the inverse calculations in the controller and estimation algorithm are still kept.

The remainder of this note is listed as following. Section II introduces the Full Form EDLM for a family of MIMO discrete-time nonlinear system. Section III presents the MFAC design and analyzes its stability. Section IV presents an example to illustrate the established results and Section V concludes the note. Finally, Appendix A gives the proof of the full-form EDLM in MIMO systems and Appendix B provides the system stability analysis in detail.

## II. EDLM FOR A FAMILY OF MULTIVARIABLE NONLINEAR SYSTEMS

This section presents a kind of full-form EDLM for a family of multivariable system, which is basic tool for MFAC design and analysis in next section.

The MIMO nonlinear system is shown as:
$$\boldsymbol{y}(i+1) = \boldsymbol{C}(\boldsymbol{y}(i),\cdots,\boldsymbol{y}(i-n_y),\boldsymbol{u}(i),\cdots,\boldsymbol{u}(i-n_u)) \quad (1)$$

where $\boldsymbol{C}(\cdots) = [C_1(\bullet),\cdots,C_m(\bullet)]^T \in \boldsymbol{R}^m$ is supposed to be the vector-valued function; $\boldsymbol{u}(i)$ and $\boldsymbol{y}(i)$ are the input vector and output vector of the system at the time of $i$, respectively; $n_y$, $n_u$ $\in R$ are the corresponding unknown orders,

We suppose that the multivariable system (1) is on the basis of the Assumptions *1* and *2*:

*Assumption 1:* The partial derivatives of $\boldsymbol{C}(\cdots)$ for all its elements are continuous.

*Assumption 2:* $C(\cdots)$ is the generalized Lipschitz function, then we have

$$|y(i_1+1) - y(i_2+1)| \le a \|L(i_1) - L(i_2)\| \quad (2)$$

where $L(i) = \begin{bmatrix} Y_{ly}(i) \\ U_{lu}(i) \end{bmatrix}$ is a matrix that contains control input vector $U_{lu}(i) = [u^T(i), \cdots, u^T(i-l_u+1)]^T$ and system output vector $Y_{ly}(i) = [y^T(i), \cdots, y^T(i-l_y+1)]^T$ with its time window $[i-l_u+1, i]$ and $[i-l_y+1, i]$, respectively. The positive integers $l_y$ and $l_u$ are named pseudo orders of the system with $1 \le l_y \le n_y$ and $1 \le l_u \le n_u$. Please refer to [1], [17] for more details about *Assumption 1* and *Assumption 2*.

*Theorem 1:* Given system (1) satisfying above two assumptions, if $\Delta L(i) \ne 0$, $1 \le l_y \le n_y$, $1 \le l_u \le n_u$, there must exist a matrix $\boldsymbol{\Phi}(i)$ named pseudo Jacobian matrix, such that (1) is expressed by the following full-form EDLM:

$$\Delta y(i+1) = \boldsymbol{\Phi}(i) \Delta L(i) \quad (3)$$

with $\|\boldsymbol{\Phi}(i)\| \le a$ for any $i$, where $\boldsymbol{\Phi}(i) = [\boldsymbol{\Phi}_1(i), \cdots, \boldsymbol{\Phi}_{Ly}(i), \boldsymbol{\Phi}_{Ly+1}(i), \cdots, \boldsymbol{\Phi}_{Ly+Lu}(i)]$,

$$\boldsymbol{\Phi}_k(i) = \begin{bmatrix} \phi_{11k}(i) & \phi_{12k}(i) & \cdots & \phi_{1mk}(i) \\ \phi_{21k}(i) & \phi_{22k}(i) & \cdots & \phi_{2mk}(i) \\ \vdots & \vdots & \vdots & \vdots \\ \phi_{m1k}(i) & \phi_{m2k}(i) & \cdots & \phi_{mmk}(i) \end{bmatrix} \in \boldsymbol{R}^{m \times m}$$

$k=1, \cdots, l_y+l_u$.

$\Delta L(i) = \begin{bmatrix} \Delta Y_{ly}(i) \\ \Delta U_{lu}(i) \end{bmatrix}$. And we suppose that $\Delta y(i) = 0$ and $\Delta u(i) = 0$ for all $i \le 0$.

*Proof:* Refer to Appendix A.

*Assumption 3:* $\boldsymbol{\Phi}_{ly+1}(i)$ is a nonsingular matrix with the sign of all the elements unchanged, and each matrix $\boldsymbol{\Phi}_k(i)$ is bounded with following sense $\|\boldsymbol{\Phi}_k(i)\| \le a$, $k=1, \cdots, m$.

*Remark 1:* For more details about FFDL modeling method, please refer to [1], [17].

## III. MFAC DESIGN WITH ITS STABILITY ANALYSIS

This section gives the controller design and stability analysis for the proposed method with some necessary Theorems and Lemmas.

### A. Model-Free Adaptive Controller Designl

We rewrite (3) into (4).

$$y(i+1) = y(i) + \boldsymbol{\Phi}(i) \Delta L(i) \quad (4)$$

The cost function is given as follow:

$$J = \left[y^d(i+1) - y(i+1)\right]^T \left[y^d(i+1) - y(i+1)\right] + \lambda \Delta u^T(i) \Delta u(i) \quad (5)$$

Where, $\lambda$ is the weighting constant; $y^d(i+1) = \left[y_1^d(i+1), \cdots, y_m^d(i+N)\right]^T$ is the set point vector.

We substitute (4) into (5) and minimize the (5) to have:

$$\Delta u(i) = [\lambda \boldsymbol{I} + \boldsymbol{\Phi}_{ly+1}^T(i) \boldsymbol{\Phi}_{ly+1}(i)]^{-1} \boldsymbol{\Phi}_{ly+1}^T(i)[(y^d(i+1) - y(i)) - \sum_{k=1}^{ly} \boldsymbol{\Phi}_k(i) \Delta y(i-k+1) - \sum_{k=ly+2}^{ly+lu} \boldsymbol{\Phi}_i(i) \Delta u(i-k+1)] \quad (6)$$

We choose one index function for unknown PJM $\boldsymbol{\Phi}(i)$ estimation as:

$$J(\boldsymbol{\Phi}(i)) = \left[\Delta y(i) - \boldsymbol{\Phi}(i) \Delta L(i-1)\right]^T \left[\Delta y(i) - \boldsymbol{\Phi}(i) \Delta L(i-1)\right] + \mu \left[\boldsymbol{\Phi}(i) - \hat{\boldsymbol{\Phi}}(i-1)\right]^T \left[\boldsymbol{\Phi}(i) - \hat{\boldsymbol{\Phi}}(i-1)\right] \quad (7)$$

Where, $\hat{\boldsymbol{\Phi}}(i)$ is the online estimated result of $\boldsymbol{\Phi}(i)$.

By the optimizing (7), the estimation algorithm for the PJM is shown as follows:

$$\hat{\boldsymbol{\Phi}}(i) = \hat{\boldsymbol{\Phi}}(i-1) + \left[\Delta y(i) - \hat{\boldsymbol{\Phi}}(i-1) \Delta L(i-1)\right]$$
$$\bullet \Delta L^T(i-1)\left[\mu \boldsymbol{I} + \Delta L(i-1)^T \Delta L(i-1)\right]^{-1} \quad (8)$$

Where, $\mu, \eta > 0$ are the adjustable parameters of estimation algorithm. The proof of blondness of PJM is similar to that in [17].

Then we introduce adjustable step factors $\rho_k < 1$ ($k=1,2,\cdots,l_y+l_u$) to have more tuning parameters and to prove the system stability. Then the controller vector is given by

$$\Delta u(i) = \hat{\boldsymbol{\xi}}_{ly+1}(i)[\rho_{ly+1}(y^d(i+1) - y(i)) - \sum_{k=1}^{ly} \rho_k \hat{\boldsymbol{\Phi}}_k(i) \Delta y(i-k+1)$$
$$- \sum_{k=ly+2}^{ly+lu} \rho_k \hat{\boldsymbol{\Phi}}_k(i) \Delta u(i-k+1)] \quad (9)$$

Where, $\hat{\boldsymbol{\xi}}_{ly+1}(i) = [\hat{\boldsymbol{\Phi}}_{ly+1}^T(i) \hat{\boldsymbol{\Phi}}_{ly+1}(i) + \lambda \boldsymbol{I}]^{-1} \hat{\boldsymbol{\Phi}}_{ly+1}^T(i)$.

### B. Stability Analysis

This section gives the proof of stability of MFAC with some Lemma and assumptions.

*Lemma 1* ([2][18]): Given $\boldsymbol{M} \in R^{n \times n}$, it has an induced consistent matrix norm such that

$$\|\boldsymbol{M}\|_v \le s(\boldsymbol{M}) + \varepsilon$$

, for any given $\varepsilon > 0$. $s(\bullet)$ represents the spectral radius of $\bullet$.

*Assumption 4:* We quote *Assumption 3* and *4* in [9] for saving room in this note.

*Theorem 2:* If the system (1) satisfies *Assumption* 1-4 and is controlled by control law (8)-(9) with the reference vector $y^d(i) = const$, there exists a $\lambda_{\min}$, when $\lambda > \lambda_{\min}$, it ensures that:

i) $\lim_{i \to \infty} |y(i+1) - y^d| = 0$; ii) $y(i)$ and $u(i)$ are bounded.

*Proof:* Inspired by [1], [9], [17], we give the proof of *Theorem 2* in Appendix B.

## IV. SIMULATIONS

Example 1: An example in [1] is selected to make comparisons between the proposed MFAC and the current MFAC, and the multivariable system cited from [1] is shown as (10).

$$\begin{cases} y_1(i+1) = \dfrac{2.5 y_1(i) y_1(i-1) + 0.09 u_1(i) u_1(i-1)}{1 + y_1^2(i) + y_1^2(i-1)} \\ \quad + 1.2 u_1(i) + 1.6 u_1(i-2) + 0.09 u_1(i) u_2(i-1) \\ \quad + 0.5 u_2(i) + 0.7 \sin(0.5(y_1(i) + y_1(i-1))) \\ \quad \cdot \cos(0.5(y_1(i) + y_1(i-1))) \\ y_2(i+1) = \dfrac{5 y_2(i) y_2(i-1)}{1 + y_2^2(i) + y_2^2(i-1) + y_2^2(i-2)} + u_2(i) \\ \quad + 1.1 u_2(i-1) + 1.4 u_2(i) + 0.5 u_1(i) \end{cases} \quad (10)$$

The given system is characterized with variable structure, discontinuous and is supposed unknown for controller design. The reference is

$$y_1^d(i+1) = 5\sin(\pi i/50) + 2\cos(\pi i/20)$$
$$y_2^d(i+1) = 2\sin(\pi i/50) + 5\cos(\pi i/20)$$

All the parameters and initial setting for both methods are all cited from [1]. The initial values are $y_1(1) = y_1(3) = y_2(1) = y_2(3) = 0$, $y_1(2) = y_2(2) = 1$, $u_1(1) = u_1(2) = u_2(1) = 1$, $u_2(2) = 0$. The controller parameters are $L_y = 1$, $L_u = 3$, $\eta = \rho_1 = \rho_2 = \rho_3 = \rho_4 = 0.5$, $\mu = 1$, $\lambda = 1$, $\hat{\boldsymbol{\Phi}}(1) = \hat{\boldsymbol{\Phi}}(2) = \begin{bmatrix} 0 & 0 & 0.1 & 0 & 0 & 0 & 0 & 0 \\ 0 & 0 & 0 & 0.1 & 0 & 0 & 0 & 0 \end{bmatrix}$. The outputs of system controlled by proposed and current MFAC are given in Fig. 1 and Fig. 2. The outputs of both controller are given in Fig. 3. In this example, we apply the same estimation algorithm in [1], so Fig. 4 only shows the time-varying parameters in $\hat{\boldsymbol{\Phi}}_{ly+1}(i)$ of the proposed MFAC for saving room. The performance indexes for both are given in Tab I.

Through Fig. 1, Fig.2 and Tab I, we can research the conclusion that the proposed MFAC controlled system has less tracking error than the current MFAC controlled one. Since the current MFAC takes the norm of the inverse matrix part of the proposed MFAC, this may lose the part of coupling relationships among inputs and outputs. To this end, we keep the inverse calculations in proposed MFAC unchanged.

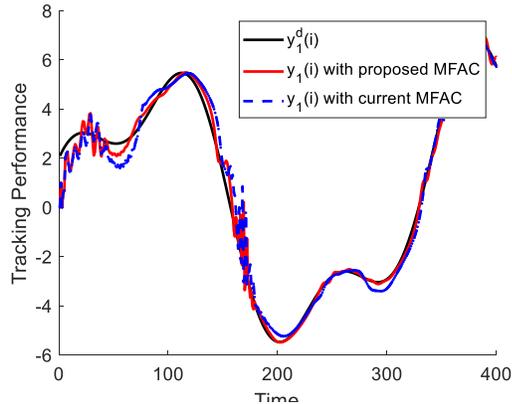

Figure. 1 Tracking performance of $y_1$

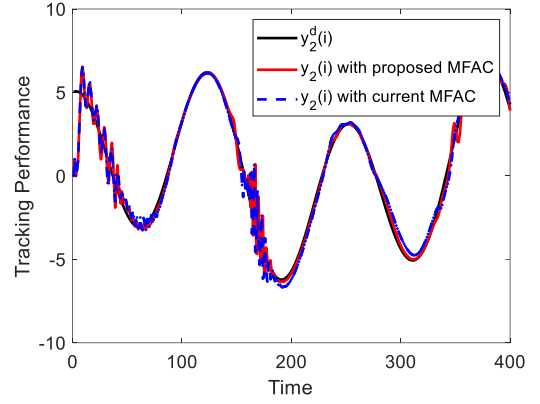

Figure. 2 Tracking performance of $y_2$

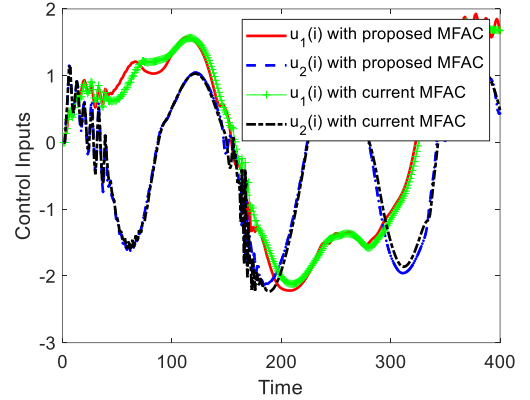

Fig. 3 Control inputs

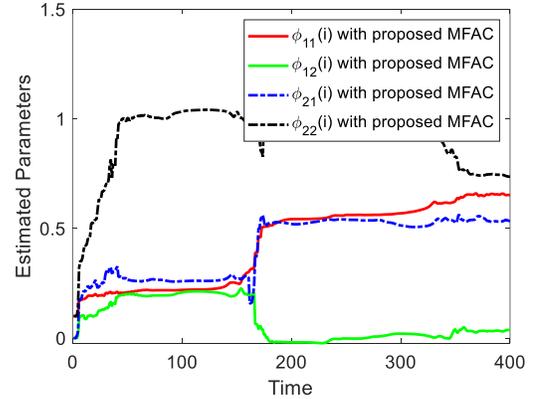

Fig. 4 Estimated parameters of PJM $\hat{\boldsymbol{\Phi}}_{Ly+1}(i)$

TABLE I Performance Indexes for both MFAC

| $eITAE = \sum_{i=1}^{N}\left| e(i) \right|^2$ | Proposed MFAC | Current MFAC |
|---|---|---|
| $y_1$ | 66.3846 | 121.8080 |
| $y_2$ | 215.0012 | 233.3453 |

## V. CONCLUSION

A kind of MFAC on the basis of full-form EDLM for a family of multivariable nonlinear systems is presented. By contraction mapping technique, we prove the tracking error astringency and analyze the BIBO stability of the multivariable system. The simulation is performed to verify the effectiveness of method.

## APPENDIX A

Proof of Theorem 1:

*Proof*:
From (1), we have
$$\Delta y(i+1) = C(y(i),\cdots,y(i-n_y),u(i),\cdots,u(i-n_u))$$
$$-C(y(i-1),\cdots,y(i-n_y-1),u(i-1),\cdots,u(i-n_u-1))$$
$$= C(y(i),\cdots,y(i-l_y+1),y(k-l_y),\cdots,y(i-n_y),$$
$$u(k),\cdots,u(k-l_u+1),u(k-l_u),\cdots,u(i-n_u))$$
$$-C(y(i-1),\cdots,y(i-l_y),y(i-l_y),\cdots,y(i-n_y),$$
$$u(i-1),\cdots,u(i-l_u),u(i-l_u),\cdots,u(i-n_u))$$
$$+C(y(i-1),\cdots,y(i-l_y),y(i-l_y),\cdots,y(i-n_y),$$
$$u(i-1),\cdots,u(i-L_u),u(i-l_u),\cdots,u(i-n_u))$$
$$-C(y(i-1),\cdots,y(i-L_y),y(i-L_y-1),\cdots,y(i-n_y-1),$$
$$u(i-1),\cdots,u(i-l_u),u(i-l_u-1),\cdots,u(i-n_u-1))$$
(11)

Let
$$\Psi(i) = C(y(i-1),\cdots,y(i-l_y),y(i-l_y),\cdots,y(i-n_y),$$
$$u(i-1),\cdots,u(i-l_u),u(i-l_u),\cdots,u(i-n_u))$$
$$-C(y(i-1),\cdots,y(i-l_y),y(i-l_y-1),\cdots,y(i-n_y-1),$$
$$u(i-1),\cdots,u(i-l_u),u(i-l_u-1),\cdots,u(i-n_u-1))$$
(12)

Due to the *Assumption 1*, (11) is transformed as (13) by the mean value theorem,
$$\Delta y(i+1) = \frac{\partial C}{\partial y(i)}\Delta y(i)+\cdots+\frac{\partial C}{\partial y(i-l_y)}\Delta y(i-l_y+1)$$
$$+\frac{\partial C}{\partial u(i)}\Delta u(i)+\cdots+\frac{\partial C}{\partial u(i-L_u)}\Delta u(i-l_u+1)+\Psi(i)$$
(13)

Where,
$$\frac{\partial C}{\partial y(i-k)} = \begin{bmatrix} \frac{\partial C_1}{\partial y_1(i-k)} & \frac{\partial C_1}{\partial y_2(i-k)} & \cdots & \frac{\partial C_1}{\partial y_m(i-k)} \\ \frac{\partial C_2}{\partial y_1(i-k)} & \frac{\partial C_2}{\partial y_2(i-k)} & \cdots & \frac{\partial C_2}{\partial y_m(i-k)} \\ \vdots & \vdots & \vdots & \vdots \\ \frac{\partial C_m}{\partial y_1(i-k)} & \frac{\partial C_m}{\partial y_2(i-k)} & \cdots & \frac{\partial C_m}{\partial y_m(i-k)} \end{bmatrix}$$
, $k=0, \cdots, l_y$-1.

$$\frac{\partial C}{\partial u(i-j)} = \begin{bmatrix} \frac{\partial C_1}{\partial u_1(i-j)} & \frac{\partial C_1}{\partial u_2(i-j)} & \cdots & \frac{\partial C_1}{\partial u_m(i-j)} \\ \frac{\partial C_2}{\partial u_1(i-j)} & \frac{\partial C_2}{\partial u_2(i-j)} & \cdots & \frac{\partial C_2}{\partial u_m(i-j)} \\ \vdots & \vdots & \vdots & \vdots \\ \frac{\partial C_m}{\partial u_1(i-j)} & \frac{\partial C_m}{\partial u_2(i-j)} & \cdots & \frac{\partial C_m}{\partial u_m(i-j)} \end{bmatrix}$$
, $j=1, \cdots, l_u$-1.

$\frac{\partial C_p}{\partial y_q(i-k)}$ ($p, q=1, \cdots, m$) and $\frac{\partial C_p}{\partial u_t(k-i)}$ ($t=1, \cdots, m$) mean the partial derivative of $C_p$ at one point in range of $[y_q(i-k), y_q(i-k-1)]$, $k=0, \cdots, l_y$-1 and $[u_t(i-j), u_t(i-j-1)]$, $j=0, \cdots, l_u$-1, respectively.

Then the below equation with variables matrix $\eta(i)_{m\times m(l_y+l_u)}$ at any time $i$ is considered
$$\Psi(i) = \eta(i)\Delta L(i) \quad (14)$$
There must exist no less than one solution $\eta^*(i)$ for equation (14), when $\Delta L(i) \neq 0$.

Let
$$\Phi(i) = \left[\frac{\partial C}{\partial y(i)},\cdots,\frac{\partial C}{\partial y(i-l_y)},\frac{\partial C}{\partial u(i)},\cdots,\frac{\partial C}{\partial u(i-l_u)}\right]+\eta(i)$$
(15)

We obtain
$$\Delta y(i+1) = \Phi(i)\Delta L(i) \quad (16)$$
$\|\Phi(i)\| \leq a$ is a direct result in *Assumption 2*.

APPENDIX B
Proof of Theorem 2:
*Proof*: This section gives proofs of the astringency of the tracking error and stability of the controlled system.

We define the vector $\Delta R(i) = \begin{bmatrix} \Delta u(i) \\ \vdots \\ \Delta u(i-l_u+1) \\ \Delta y(i) \\ \vdots \\ \Delta y(i-l_y+1) \end{bmatrix}_{m(Ly+Lu)\times m}$, and the

matrixes
$$\hat{\Psi}_U(i) = [\hat{\Phi}_{ly+2}(i),\hat{\Phi}_{ly+3}(i),\cdots,\hat{\Phi}_{ly+lu}(i),\mathbf{0}]_{m\times m\bullet lu},$$
$$\hat{\Psi}_Y(i) = [\hat{\Phi}_1(i),\hat{\Phi}_2(i),\cdots,\hat{\Phi}_{Ly}(i)]_{m\times m\bullet ly}.$$

Then we have
$$\Delta \gamma(i)$$
$$=[\Delta u^T(i),\cdots,\Delta u^T(i-l_u+1),\Delta y^T(i),\cdots,\Delta y^T(i-l_y+1)]^T$$
$$= \begin{bmatrix} \hat{\xi}_{ly+1}(i)[\rho_{Ly+1}(y^d(i+1)-y(i))-\sum_{k=1}^{ly}\rho_k\hat{\Phi}_k(i)\Delta y(i-k+1) \\ -\sum_{k=ly+2}^{ly+lu}\rho_k\hat{\Phi}_k(i)\Delta u(i-k+1)] \\ \vdots \\ \Delta u(i-l_u+1) \\ \Delta y(i) \\ \vdots \\ \Delta y(i-l_y+1) \end{bmatrix}$$
$$=A(i)\gamma_1 + \rho_{ly+1}E\hat{\xi}_{ly+1}(i)e(i)$$
(17)

where $\gamma_1 = [\Delta u^T(i-1),\cdots,\Delta u^T(i-l_u+1),\Delta y^T(i),\cdots,$
$\Delta y^T(i-l_y+1),\Delta u^T(i-l_u)]^T$
$= FD(i-1)\Delta R(i-1)$

$$A(i)_{m(ly+lu)\times m(ly+lu)} =$$

$$\begin{bmatrix} -p_{ly+2} & \cdots & -p_{ly+lu} & -p_1 & \cdots & -p_{ly} & 0 \\ I & & & & & & \\ 0 & \ddots & 0 & & & & \\ & & I & \ddots & & & \\ & & & \ddots & 0 & & \\ & & & & \ddots & \ddots & \\ & & & & & I & 0 \end{bmatrix}$$

$p_k = \rho_k \hat{\xi}_{ly+1}(i) \hat{\boldsymbol{\Phi}}_k(i)$, $k = 1,\cdots, l_y, l_y+2,\cdots, l_y+l_u$.

$$F = \begin{bmatrix} I & & & & & & & \\ & \ddots & & & & & & \\ & & I & & & & & \\ & & & 0 & I & & & \\ & & & I & & & & \\ & & & & & \ddots & 0 & \\ & & & I & 0 & & 0 \end{bmatrix}_{m(ly+lu)\times m(ly+lu)}$$

$$E = \begin{bmatrix} I_{m\times m} \\ 0_{m(ly+lu-1)\times m} \end{bmatrix}$$

$$D(i-1) = \begin{bmatrix} I & & & & & & & \\ & \ddots & & & & & & \\ & & I & & & & & \\ & & & I & & & & \\ & & & & \ddots & & & \\ & & & & & I & & \\ \boldsymbol{\Phi}_{ly+1} & \cdots & \boldsymbol{\Phi}_{ly+lu} & \boldsymbol{\Phi}_1 & \cdots & \boldsymbol{\Phi}_{ly-1} & \boldsymbol{\Phi}_{ly} \end{bmatrix}_{m(ly+lu)\times m(ly+lu)}$$

(17) may be expressed as

$$\Delta \gamma(i) = A(i) F D(i-1) \Delta \gamma(i-1) + \rho_{ly+1} E \hat{\xi}_{ly+1}(i) e(i) \quad (18)$$

Considering the following operation for the first $m$ rows in $A(i)$ and according to the matrix norm inequalities about the relationship between $\|\bullet\|_\infty$ and $\|\bullet\|_2$ in [18], we can get

$$\sum_{k=ly+2}^{ly+lu} \left\| \rho_{ly+k+1} \hat{\xi}_{ly+1} \hat{\boldsymbol{\Phi}}_k(i) \right\|_\infty + \sum_{k=1}^{ly} \left\| \rho_{lk} \hat{\xi}_{ly+1} \hat{\boldsymbol{\Phi}}_k(i) \right\|_\infty$$
$$\leq (\max_{\substack{k=1,\cdots,ly,ly+2, \\ \cdots, ly+lu}} \rho_k) \left\| [\hat{\boldsymbol{\Phi}}_{ly+1}^T(i)\hat{\boldsymbol{\Phi}}_{ly+1}(i) + \lambda I]^{-1} \hat{\boldsymbol{\Phi}}_{ly+1}^T(i) \right\|_\infty$$
$$\bullet \sum_{k=1, i\neq ly+1}^{ly+lu} \left\| \hat{\boldsymbol{\Phi}}_k(i) \right\|_\infty \quad (19)$$
$$\leq (\max_{\substack{k=1,\cdots,ly,ly+2, \\ \cdots, ly+lu}} \rho_k) \sqrt{m} \left\| [\hat{\boldsymbol{\Phi}}_{ly+1}^T(i)\hat{\boldsymbol{\Phi}}_{ly+1}(i) + \lambda I]^{-1} \right\|_2 \left\| \hat{\boldsymbol{\Phi}}_{ly+1}^T(i) \right\|_\infty$$
$$\bullet \sum_{k=1, k\neq ly+1}^{ly+lu} \left\| \hat{\boldsymbol{\Phi}}_k(i) \right\|_\infty$$

Since $\hat{\boldsymbol{\Phi}}_{ly+1}^T(i)\hat{\boldsymbol{\Phi}}_{ly+1}(i)$ is a symmetric semi-positive matrix, $\hat{\boldsymbol{\Phi}}_{ly+1}^T(i)\hat{\boldsymbol{\Phi}}_{ly+1}(i) + \lambda I$ is also a symmetric positive matrix and we can obtain that $\left[ \lambda I + \left(\hat{\boldsymbol{\Phi}}_{ly+1}^T(i)\hat{\boldsymbol{\Phi}}_{ly+1}(i)\right)^{-1} \right]^T = \left( \lambda I + \hat{\boldsymbol{\Phi}}_{ly+1}^T(i)\hat{\boldsymbol{\Phi}}_{ly+1}(i) \right)^{-1}$. Where, $\|\bullet\|_\infty$ represents the maximum row sum matrix norm. $\|\bullet\|_2$ represents the spectral norm of matrix. Given the eigenvalues of $\hat{\boldsymbol{\Phi}}_{ly+1}^T(i)\hat{\boldsymbol{\Phi}}_{ly+1}(i)$ are $s_k \geq 0$, $k = 1,\cdots, N_u$, the corresponding eigenvalues of $\hat{\boldsymbol{\Phi}}_{ly+1}^T(i)\hat{\boldsymbol{\Phi}}_{ly+1}(i) + \lambda I$ will be $\lambda + s_k > 0$, which implies the eigenvalues of $[\hat{\boldsymbol{\Phi}}_{ly+1}^T(i)\hat{\boldsymbol{\Phi}}_{ly+1}(i) + \lambda I]^{-1}$ will be $\frac{1}{\lambda + s_k} > 0$. Therefore, we obtain

$$\left\| [\hat{\boldsymbol{\Phi}}_{ly+1}^T(i)\hat{\boldsymbol{\Phi}}_{ly+1}(i) + \lambda I]^{-1} \right\|_2$$
$$= \sqrt{\sigma\left( \left([\lambda I + \hat{\boldsymbol{\Phi}}_{ly+1}^T(i)\hat{\boldsymbol{\Phi}}_{ly+1}(i)]^{-1}\right)^T [\lambda I + \hat{\boldsymbol{\Phi}}_{ly+1}^T(i)\hat{\boldsymbol{\Phi}}_{ly+1}(i)]^{-1} \right)}$$
$$= \sqrt{\sigma\left( \left([\lambda I + \hat{\boldsymbol{\Phi}}_{ly+1}^T(i)\hat{\boldsymbol{\Phi}}_{ly+1}(i)]^{-1}\right)^2 \right)} = \frac{1}{\min_{k=1,\cdots Nu}\{\lambda + s_k\}}$$
(20)

Combining (19) and (20), we have

$$\sum_{k=ly+2}^{ly+lu} \left\| \hat{\xi}_{ly+1} \hat{\boldsymbol{\Phi}}_k(i) \right\|_\infty + \sum_{k=1}^{ly} \left\| \hat{\xi}_{ly+1} \hat{\boldsymbol{\Phi}}_k(i) \right\|_\infty$$
$$\leq \left\| [\lambda I + \hat{\boldsymbol{\Phi}}_{ly+1}^T(i)\hat{\boldsymbol{\Phi}}_{ly+1}(i)]^{-1} \hat{\boldsymbol{\Phi}}_{ly+1}^T(i) \sum_{k=1,k\neq ly+1}^{ly+lu} \hat{\boldsymbol{\Phi}}_k(i) \right\|_\infty$$
$$\leq \sqrt{m} \left\| [\hat{\boldsymbol{\Phi}}_{ly+1}^T(i)\hat{\boldsymbol{\Phi}}_{ly+1}(i) + \lambda I]^{-1} \right\|_2 \left\| \hat{\boldsymbol{\Phi}}_{ly+1}^T(i) \right\|_\infty \sum_{k=1,k\neq ly+1}^{ly+lu} \left\| \hat{\boldsymbol{\Phi}}_k(i) \right\|_\infty$$
$$\leq \sqrt{m} \frac{1}{\min_{k=1,\cdots Nu}\{\lambda + s_k\}} \left\| \hat{\boldsymbol{\Phi}}_{ly+1}^T(i) \right\|_\infty \sum_{k=1,k\neq ly+1}^{ly+lu} \left\| \hat{\boldsymbol{\Phi}}_k(i) \right\|_\infty$$
(21)

According to *Assumption 3*, we have $\left\| \hat{\boldsymbol{\Phi}}_{ly+1}(i) \right\|_\infty < a_1$ and $\sum_{k=1,k\neq ly+1}^{ly+lu} \left\| \hat{\boldsymbol{\Phi}}_k(i) \right\|_\infty \leq 2(l_y + l_u - 1)a_1$. Consequently, there is a positive $\lambda_{\min 1}$, if $\lambda > \lambda_{\min 1}$, it yields the following inequation:

$$\left[ \sum_{k=1}^{ly} \left\| \hat{\xi}_{ly+1} \hat{\boldsymbol{\Phi}}_k(i) \right\|_\infty + \sum_{k=ly+2}^{ly+lu} \left\| \hat{\xi}_{ly+1} \hat{\boldsymbol{\Phi}}_k(i) \right\|_\infty \right]^{\frac{1}{ly+lu-1}}$$
$$\leq \left[ \sqrt{m} \frac{1}{\min_{k=1,\cdots Nu}\{\lambda + b_k\}} \left\| \hat{\boldsymbol{\Phi}}_{ly+1}^T(i) \right\|_\infty \bullet \sum_{k=1,k\neq ly+1}^{ly+lu} \left\| \hat{\boldsymbol{\Phi}}_k(i) \right\|_\infty \right]^{\frac{1}{ly+lu-1}} \quad (22)$$
$$\leq M_1 < 1$$

Given $0 < \rho_1 < 1$, $\cdots$, $0 < \rho_{ly} < 1$, $0 < \rho_{ly+2} < 1$, $\cdots$, $0 < \rho_{ly+lu} < 1$, it has $(\max_{\substack{k=1,\cdots,ly,ly+2, \\ \cdots, ly+lu}} \rho_k) < 1$. The we obtain

$$\sum_{k=1}^{lu-1} \left\| \rho_{ly+k+1} \hat{\xi}_{ly+1}(i) \hat{\boldsymbol{\Phi}}_k(i) \right\|_\infty + \sum_{k=1}^{ly} \left\| \rho_k \hat{\xi}_{ly+1}(i) \hat{\boldsymbol{\Phi}}_k(i) \right\|_\infty$$
$$\leq (\max_{\substack{k=1,\cdots,ly,ly+2, \\ \cdots, ly+lu}} \rho_k) M_1^{ly+lu-1} < 1 \quad (23)$$

According to [1] and [17], we will have the characteristic equation of $A(k)$:

$$z^{m(ly+lu-1)}(z^{ly+lu}I + \rho_{ly+2}\hat{\xi}_{ly+1}(i)\hat{\Phi}_{ly+2}(i)z^{ly+lu-1} + \cdots \\ + \rho_1\hat{\xi}_{ly+1}(i)\hat{\Phi}_1(i)z^{ly} + \cdots + \rho_{ly}\hat{\xi}_{ly+1}(i)\hat{\Phi}_{ly}(i)z = 0 \quad (24)$$

Based on (24), we have (25).

$$|z|^{ly+lu-1} \leq (\max_{\substack{k=1,\cdots,ly,ly+2,\\ \cdots,ly+lu}} \rho_i)\left[\sum_{k=1}^{lu-1}\left\|\hat{\xi}_{ly+1}\hat{\Phi}_{ly+1+k}(i)\right\|_\infty + \sum_{k=1}^{ly}\left\|\hat{\xi}_{ly+1}\hat{\Phi}_k(i)\right\|_\infty\right] \\ \leq (\max_{\substack{i=1,\cdots,ly,ly+2,\\ \cdots,ly+lu}} \rho_i)M_1^{ly+lu-1} < 1 \quad (25)$$

which means $|z| \leq (\max_{\substack{i=1,\cdots,ly,ly+2,\\ \cdots,ly+lu}} \rho_i)^{1/ly+lu-1}M_1 < 1$. In light of *Lemma 1* and (25), there will be an arbitrarily small positive ε and a compatible norm $\|\bullet\|_v$ obtaining the following inequation.

$$\|A(i)\|_v \leq s(A(i)) + \varepsilon \leq (\max_{i=1,\cdots,ly+lu}\rho_i)^{1/ly+lu-1}M_1 + \varepsilon < 1 \quad (26)$$

where $\|\bullet\|_v$ represents the compatible norm of $\bullet$. Let $d_1 = (\max_{\substack{k=1,\cdots,ly,ly+2,\\ \cdots,ly+lu}} \rho_k)^{1/ly+lu-1}M_1$.

By the definition of spectral radius, [9] has achieved the following inequation

$$\|A(i)\|_v\|F\|_v\|D(i-1)\|_v \\ \leq (d_1+\varepsilon)(1+\varepsilon)(\max(1,b)+\varepsilon) \triangleq d_2 < 1 \quad (27)$$

Samilar to (21), we have

$$\left\|\hat{\xi}_{ly+1}\right\|_\infty = \left\|[\lambda I + \Phi_{ly+1}^T(i)\Phi_{ly+1}(i)]^{-1}\Phi_{ly+1}^T(i)\right\|_\infty \\ \leq \sqrt{m}\left\|[\lambda I + \hat{\Psi}_{Nu}^T(i)\hat{\Psi}_{Nu}(i)]^{-1}\right\|_2\left\|\Phi_{ly+1}^T(i)\right\|_\infty \\ \leq \sqrt{m}\frac{1}{\min_{k=1,\cdots Nu}\{\lambda+b_k\}}\left\|\Phi_{ly+1}^T(i)\right\|_\infty \quad (28)$$

Therefore, it has positive $\lambda_{\min 2}$ and $M_2$, when $\lambda > \lambda_{\min 2}$, we obtain (29) and (30).

$$0 < \left\|\hat{\xi}_{ly+1}(i)\right\|_v \leq b\left\|\hat{\xi}_{ly+1}(i)\right\|_\infty \leq M_2 < 1 \quad (29)$$

$$d_3 \triangleq \rho_{ly+1}M_2\left\|X\phi_{ly}^T(i+1)\right\|_v < 0.5 \quad (30)$$

where, $X = \begin{bmatrix} I_{lu} \\ & I_{ly} \end{bmatrix}$, $b$ is a constant according to the equivalence theorem of matrix norm [18].

We combine (27), (29) and the norm of (17) together to obtain

$$\|\Delta\gamma(i)\|_v = \|A(i)\|_v\|F\|_v\|D\|_v\|\Delta\gamma(i-1)\|_v + \rho_{ly+1}\left\|E\hat{\xi}_{ly+1}(i)\right\|_v\|e(i)\|_v \\ = d_2\|\Delta\gamma(i-1)\|_v + \rho_{ly+1}M_2\|e(i)\|_v \\ \vdots \\ = d_2^i\|\Delta\gamma(0)\|_v + \rho_{ly+1}M_2\sum_{k=1}^i d_2^{i-k}\|e(k)\|_v \\ < \rho_{ly+1}M_2\sum_{k=1}^i d_2^{i-k}\|e(k)\|_v \quad (31)$$

We combine (3) and (18) together to have

$$e(i+1) = y^d - y(i+1) = y^d - y(i) - \Phi(i)\Delta L(i) \\ = e(i) - X\Phi(i)[A(k)FD(i-1)\Delta\gamma(i-1) + \rho_{ly+1}E\hat{\xi}_{ly+1}(i)e(i)] \\ = \left[I - \rho_{ly+1}\Phi_{ly+1}(i)\hat{\xi}_{ly+1}(i)\right]e(i) - X\Phi(i)A(i)FD\Delta\gamma(i-1) \quad (32)$$

Since

$$[\hat{\Phi}_{ly+1}^T(i)\hat{\Phi}_{ly+1}(i) + \lambda I]^{-1}\hat{\Phi}_{Ly+1}^T(i) \\ = \frac{1}{\lambda}[\hat{\Phi}_{ly+1}^T(i)\frac{\hat{\Phi}_{ly+1}(i)}{\lambda} + I]^{-1}\hat{\Phi}_{ly+1}^T(i) \\ = \hat{\Phi}_{ly+1}^T(i)[\hat{\Phi}_{ly+1}(i)\hat{\Phi}_{ly+1}^T(i) + \lambda I]^{-1} \quad (33)$$

then we have

$$\left\|I - \rho_{ly+1}\Phi_{ly+1}(i)\hat{\xi}_{ly+1}\right\|_\infty \\ = \left\|I - \rho_{ly+1}\Phi_{ly+1}(i)\hat{\Phi}_{ly+1}^T(i)[\hat{\Phi}_{ly+1}(i)\hat{\Phi}_{ly+1}^T(i) + \lambda I]^{-1}\right\|_\infty \\ = \left\|I - \rho_{ly+1}\Phi_{ly+1}(i)\hat{\Phi}_{ly+1}^T(i)[\hat{\Phi}_{ly+1}(i)\hat{\Phi}_{ly+1}^T(i) + \lambda I]^{-1}\right\|_\infty \\ = \left\|[\hat{\Phi}_{ly+1}(i)\hat{\Phi}_{ly+1}^T(i) - \rho_{ly+1}\Phi_{ly+1}(i)\hat{\Phi}_{ly+1}^T(i) + \lambda I] \\ \bullet[\hat{\Phi}_{ly+1}(i)\hat{\Phi}_{ly+1}^T(i) + \lambda I]^{-1}\right\|_\infty \triangleq d_4 \quad (34)$$

Similarly, there exists the adjustable parameter $\rho_{ly+1}$ and positive $\lambda_{\min 3}$, such that $\lambda > \lambda_{\min 3}$, then we have the following inequation.

$$d_3 < 0.5 < d_4 = \left\|I - \rho_{ly+1}\Phi_{ly+1}(i)\hat{\xi}_{ly+1}\right\|_\infty < M_3 < 1 \quad (35)$$

We take the norm of (32) and use the $\|\Delta R(0)\|_v = 0$ (*Theorem 1*) to get

$$\|e(i+1)\|_v = \left\|I - \rho_{ly+1}\Phi_{ly+1}(i)E(i)\hat{\xi}_{ly+1}(i)\right\|_v\|e(i)\|_v \\ -\|\Phi_{ly+1}(i)\|_v\|A(i)FD(i-1)\|_v\|\Delta\gamma(i-1)\|_v \\ < d_4\|e(i)\|_v + d_2\|\Phi_{ly+1}(i)\|_v\|\Delta\gamma(i-1)\|_v < \cdots \\ < d_4^{i-1}\|e(i)\|_v + d_2\sum_{k=1}^{i-1}d_4^{i-1-k}\|\Phi_{ly+1}(i+1)\|_v\|\Delta\gamma(i-1)\|_v \\ < d_4^{i-1}\|e(2)\|_v \\ + d_2\sum_{k=1}^{i-1}d_4^{i-1-k}\|\Phi_{ly+1}(k+1)\|_v[\rho_{ly+1}M_2\sum_{j=1}^k d_2^{k-j}\|e(j)\|_v] \quad (36)$$

Then (36) becomes

$$\|e(i+1)\|_v < d_4^{i-1}\|e(2)\|_v + d_2d_3\sum_{k=1}^{i-1}d_4^{i-1-k}\sum_{j=1}^k d_2^{k-j}\|e(j)\|_v \quad (37)$$

Let

$$g(i+1) = d_4^{i-1}\|e(2)\|_v + d_2d_3\sum_{k=1}^{i-1}d_4^{i-1-k}\sum_{j=1}^k d_2^{k-j}\|e(j)\|_v \quad (38)$$

Then inequation (37) may be rewritten as below

$$\|e(i+1)\|_v < g(i+1), \quad i=1, 2 \ldots \quad (39)$$

where, $g(2) = d_3\|e(1)\|_v$. According to (38) and (39), we have

$$g(i+2) = d_4^i \|e(2)\|_v + d_2 d_3 \sum_{k=1}^{i} d_4^{i-k} \sum_{j=1}^{k} d_2^{k-j} \|e(j)\|_v$$

$$= d_4 g(i+1) + d_2 d_3 \sum_{j=1}^{i-1} d_2^{i-j} \|e(j)\|_v + d_2 d_3 \|e(i)\|_v \quad (40)$$

$$\leq d_4 g(i+1) + d_2 d_3 \sum_{j=1}^{i-1} d_2^{i-j} \|e(j)\|_v + d_2 d_3 g(i)$$

Let

$$h(i) = d_2 d_3 \sum_{j=1}^{i-1} d_2^{i-j} \|e(j)\|_v + d_2 d_3 g(i) \quad (41)$$

Then (41) combines with (35), we have

$$h(i) < d_2 d_3 \sum_{j=1}^{i-1} d_2^{i-j} \|e(j)\|_v + d_2 d_4 g(i)$$

$$= d_2 d_3 \sum_{j=1}^{i-1} d_2^{i-j} \|e(j)\|_v + d_2 d_4 [d_4^{i-2} \|e(2)\|_v \quad (42)$$

$$+ d_2 d_3 \sum_{k=1}^{i-2} d_4^{i-2-k} \sum_{j=1}^{k} d_2^{k-j} \|e(j)\|_v ]$$

$$= d_2 g(i+1)$$

We substitute (42) into (40) to get

$$g(i+2) < (d_4 + d_2) g(i+1) \quad (43)$$

On the basis of $0 < \rho_k < 1$, $(k = 1, 2, \cdots, l_y, l_y+2, \cdots, l_y+l_u)$, we can get $0 < (\max_{\substack{k=1,\cdots,l_y,l_y+2,\\ \cdots,l_y+l_u}} \rho_i)^{1/l_y+l_u-1} M_1 < M_3 < 1$ to have the following inequation

$$d_4 + d_2 = 1 - M_3 + (\max_{k=1,\cdots,l_y+l_u} \rho_k)^{1/l_y+l_u-1} M_1 < 1 \quad (44)$$

At last, we have the following inequation by substituting (44) into (43).

$$\lim_{i \to \infty} g(i+2) < \lim_{i \to \infty} (d_4+d_2) g(i+1) < \cdots < \lim_{i \to \infty} (d_4+d_2)^i g(2) = 0 \quad (45)$$

*Theorem 2* is the direct conclusion of (45) and (39) under the condition that $\lambda > \lambda_{min} = \max\{\lambda_{min1}, \lambda_{min2}, \lambda_{min3}\}$.

Since $\gamma(k)$ is a vector which only contains the incremental inputs and incremental outputs of system. Therefore, the BIBO stability of the system can be achieved by proving $\gamma(k)$ is bounded.

Combine (31), (39) and (44) together, then we will have

$$\|\gamma(i)\|_v \leq \sum_{k=0}^{i} \|\Delta\gamma(k)\|_v \leq \sum_{j=1}^{i} \rho_{l_y+1} M_2 \sum_{k=1}^{j} d_2^{j-k} \|e(k)\|_v$$

$$< \frac{\rho_{l_y+1} M_2}{1-d_2} \sum_{j=1}^{i} \|e(j)\|_v < \frac{\rho_{l_y+1} M_2}{1-d_2} \sum_{j=1}^{i} g(j) \quad (46)$$

$$< \frac{\rho_{l_y+1} M_2 |g(2)|}{(1-d_2)(1-d_2-d_4)}$$

The blondness of $\|\gamma(i)\|_v$ is the direct result of (46). It also shows that the system controlled by MFAC is guaranteed BIBO stability. *Theorem* 2 is proved.